\pdfoutput=1
\documentclass[preprint,preprintnumbers,amsmath,amssymb,floatfix,endfloats*]{revtex4}

\usepackage{graphicx}
\usepackage{dcolumn}
\usepackage{bm}
\usepackage{amsfonts}

\DeclareMathAlphabet{\mathsfsl}{OT1}{cmr}{bx}{it}
\begin{document}
\title{The yielding transition in periodically sheared binary glasses at finite temperature}
\author{Nikolai V. Priezjev$^{1,2}$}
\affiliation{$^{1}$Department of Mechanical and Materials
Engineering, Wright State University, Dayton, OH 45435}
\affiliation{$^{2}$National Research University Higher School of
Economics, Moscow 101000, Russia}
\date{\today}
\begin{abstract}

Non-equilibrium molecular dynamics simulations are performed to
investigate the dynamic behavior of three-dimensional binary glasses
prepared via an instantaneous quench across the glass transition. We
found that with increasing strain amplitude up to a critical value,
the potential energy approaches lower minima in steady state,
whereas the amplitude of shear stress oscillations becomes larger.
Below the yielding transition, the storage modulus dominates the
mechanical response, and the gradual decay of the potential energy
over consecutive cycles is accompanied by reduction in size of
transient clusters of atoms with large nonaffine displacements. In
contrast, above the yield strain, the loss modulus increases and the
system settles at a higher level of potential energy due to
formation of a system-spanning shear band after a number of
transient cycles.

\vskip 0.5in


\end{abstract}

\maketitle

\section{Introduction}

The deformation and flow dynamics of yield stress materials, which
include foams, gels and glasses, generally depend on physical aging
processes, shear-induced rejuvenation, and shear banding, as well as
wall slip and finite-size effects~\cite{Manneville17}.  At the
microscopic scale, the elementary plastic deformation in disordered
materials occurs in a small volume occupied by a group of particles
undergoing irreversible rearrangements, sometimes referred to as the
shear transformation zone~\cite{Argon79,Spaepen77,Falk98}. Thus, it
was recently shown that during tension-compression cyclic loading of
metallic glasses, the initiation of a shear band takes place at the
sample surface when aggregates of shear transformation zones reach a
critical size~\cite{GaoNano15}. In addition, the results of cyclic
nanoindentation tests on metallic glasses have revealed that
hardening occurs due to irreversible particle displacements in small
volumes beneath the indenter and stiffening of the preferred
yielding path~\cite{Schuh08,Schuh12,LiJNCS15,WangASS17,Lashgari18}.
Despite significant experimental and computational efforts, the
combined effect of microplasticity and confined geometry on the
mechanical properties of disordered solids remains not fully
understood.

\vskip 0.05in

In the last few years, the mechanical response of disordered solids
to oscillatory shear deformation was investigated by a number of
groups using molecular dynamics
simulations~\cite{Priezjev13,Sastry13,Reichhardt13,Priezjev14,IdoNature15,Priezjev16,Priezjev16a,Kawasaki16,Priezjev17,Sastry17,OHern17,Hecke17,Keblinsk17,Priezjev18}.
It was found that in the elastic range of deformation at zero
temperature, particles with large amplitudes of repetitive
displacements are organized into clusters~\cite{IdoNature15}, while
at finite temperatures some trajectories become
irreversible~\cite{Priezjev13,Priezjev16,Priezjev16a}.  When the
thermal fluctuations are not important, the yielding transition is
accompanied by a sharp increase of the irreversible particle
diffusion, whereas the static structure remains
unaffected~\cite{Kawasaki16}.   Interestingly, it was recently shown
for five model glasses that the loss modulus exhibits a
characteristic peak in the high-frequency regime that overlaps with
the range of natural vibrational frequencies, while at low
frequencies, persistent damping arises from long time-scale local,
irreversible deformation~\cite{Keblinsk17}.    Furthermore, cyclic
loading with the strain amplitude below (above) the yield point
results in the formation of transient clusters (a permanent shear
band) of atoms with large nonaffine displacements~\cite{Priezjev17}.
With increasing strain amplitude, the thickness of the shear band
increases until it becomes comparable to the linear system
size~\cite{Priezjev17,Sastry17}. However, the exact mechanism of
shear band formation in periodically driven disordered solids has
not yet been determined.

\vskip 0.05in

In the recent study, the periodic deformation of binary glasses,
prepared via instantaneous quench from a liquid phase to
temperatures of about two to six orders of magnitude smaller than
the glass transition temperature, was examined using atomistic
simulations~\cite{Priezjev18}. In particular, it was shown that the
number of cycles required to reach a state with minimum potential
energy is larger at higher temperatures and/or larger strain
amplitudes. Moreover, the gradual decrease in potential energy was
found to correlate with the size of clusters of atoms with large
nonaffine displacements.  On the other hand, the amplitude of shear
stress oscillations in the elastic regime reaches a maximum value
when a large part of the system starts to deform
reversibly~\cite{Priezjev18}.

\vskip 0.05in

In this paper, we use non-equilibrium molecular dynamics simulations
to study the dynamic response of poorly annealed binary glasses to
periodic shear at a temperature approximately one quarter of the
glass transition temperature.  In agreement with the results of the
previous study~\cite{Priezjev18}, we find that below the yielding
transition, the potential energy gradually decreases to a level that
is deeper for larger strain amplitudes.  Above the yield strain, the
potential energy approaches a higher value, and the amplitude of
shear stress oscillations is reduced in the steady state, which is
characterized by a system-spanning shear band and enhanced diffusion
of particles. It will be shown that the transition from transient
clusters to a permanent shear band is reflected in the shape of the
probability distribution function of nonaffine displacements.

\vskip 0.05in

The remainder of the paper is organized as follows. In the next
section, the molecular dynamics simulation method is described. The
simulation results for the potential energy, shear stress,
mechanical properties, and nonaffine displacements of atoms as a
function of the strain amplitude are presented in
Sec.\,\ref{sec:Results}. A brief summary of the results is given in
the last section.

\section{Molecular dynamics simulations}
\label{sec:MD_Model}

The model glass is represented by the three-dimensional (80:20)
binary mixture originally introduced by Kob and
Andersen~\cite{KobAnd95} to describe the amorphous metal alloy
$\text{Ni}_{80}\text{P}_{20}$~\cite{Weber85}.  In the Kob-Andersen
(KA) model, the interaction between neighboring atoms of types
$\alpha,\beta=A,B$ is specified via the truncated Lennard-Jones (LJ)
potential:
\begin{equation}
V_{\alpha\beta}(r)=4\,\varepsilon_{\alpha\beta}\,\Big[\Big(\frac{\sigma_{\alpha\beta}}{r}\Big)^{12}\!-
\Big(\frac{\sigma_{\alpha\beta}}{r}\Big)^{6}\,\Big],
\label{Eq:LJ_KA}
\end{equation}
with the following parametrization $\varepsilon_{AA}=1.0$,
$\varepsilon_{AB}=1.5$, $\varepsilon_{BB}=0.5$, $\sigma_{AB}=0.8$,
$\sigma_{BB}=0.88$, and $m_{A}=m_{B}$~\cite{KobAnd95}. This choice
of interaction parameters prevents crystallization below the glass
transition temperature~\cite{KobAnd95}.   The cutoff radius is fixed
$r_{c,\,\alpha\beta}=2.5\,\sigma_{\alpha\beta}$ to reduce
computational cost. All results are reported in the reduced LJ units
of length, mass, energy, and time, which are set to
$\sigma=\sigma_{AA}$, $m=m_{A}$, $\varepsilon=\varepsilon_{AA}$, and
$\tau=\sigma\sqrt{m/\varepsilon}$, respectively.   The integration
of the equations of motion was performed using the velocity Verlet
algorithm~\cite{Allen87} with the time step $\triangle
t_{MD}=0.005\,\tau$~\cite{Lammps}.

\vskip 0.05in

\begin{figure}[t]
\includegraphics[width=9.0cm,angle=0]{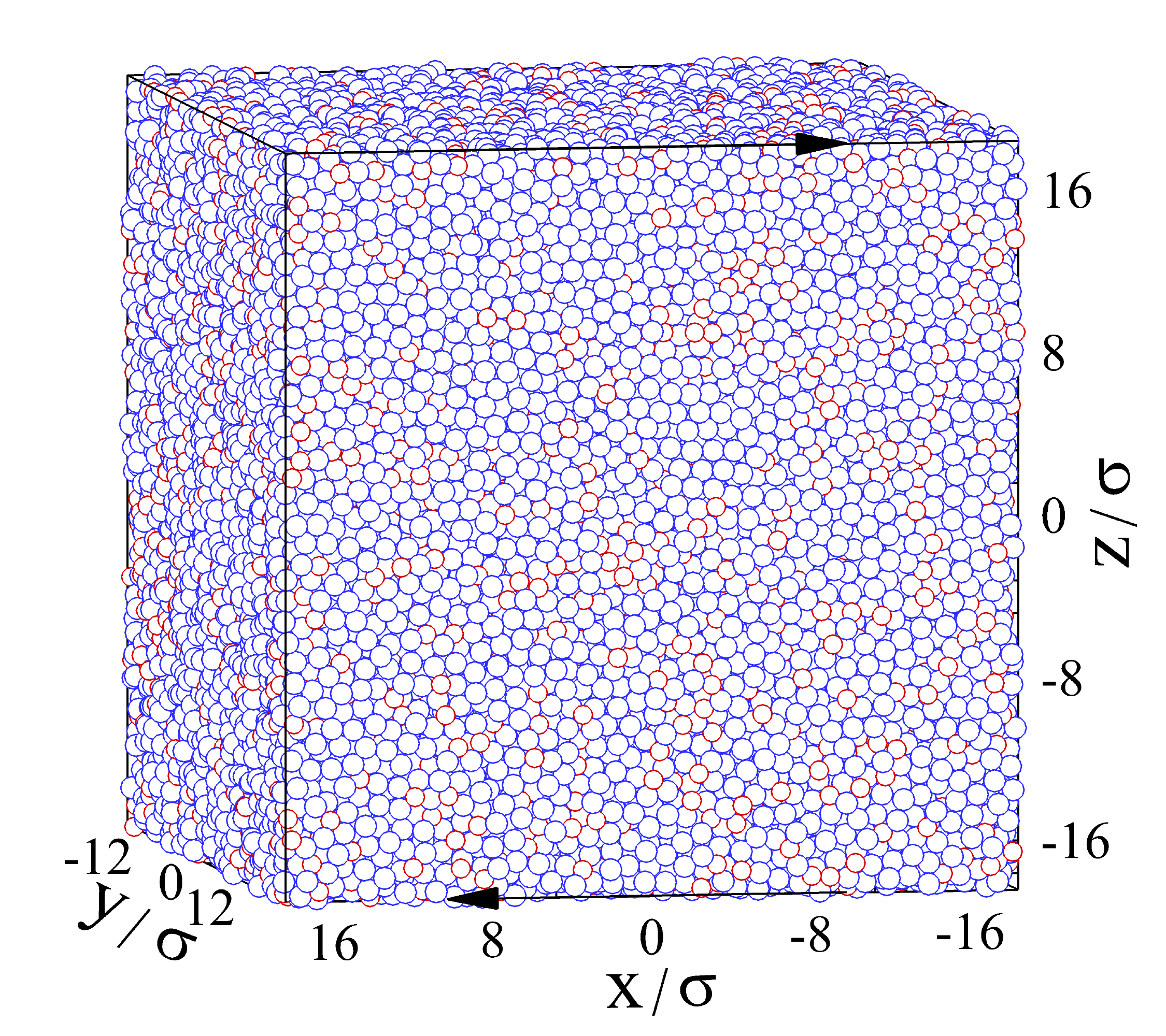}
\caption{(Color online) A snapshot of the annealed Lennard-Jones
binary glass after 600 shear cycles with the strain amplitude
$\gamma_0=0.05$. The temperature is $T_{LJ}=0.1\,\varepsilon/k_B$
and the oscillation period is $T=5000\,\tau$.   The plane of shear
is denoted by black arrows.   Atoms of types $A$ and $B$ (blue and
red circles) are now drawn to scale. }
\label{fig:snapshot_system}
\end{figure}

The equilibration was first performed at the high temperature of
$1.1\,\varepsilon/k_B$, which is well above the critical temperature
$T_c\approx0.435\,\varepsilon/k_B$ of the KA model~\cite{KobAnd95}.
Here, $k_B$ denotes the Boltzmann constant.  All simulations were
carried out at a constant volume and the atomic density
$\rho=\rho_{A}+\rho_{B}=1.2\,\sigma^{-3}$.  The total number of
atoms is $N=60\,000$ and the linear size of the periodic cubic cell
is $L=36.84\,\sigma$.  Next, following an instantaneous quench
across the glass transition to the temperature
$T_{LJ}=0.1\,\varepsilon/k_B$, the system was subjected to time
periodic shear deformation as follows:
\begin{equation}
\gamma(t)=\gamma_{0}\,\,\textrm{sin}(2\pi t / T),
\label{Eq:strain}
\end{equation}
where $\gamma_{0}$ is the strain amplitude and $T$ is the period of
oscillation.   The shear deformation was applied parallel to the
$xz$ plane (see Fig.\,\ref{fig:snapshot_system}) by using the
Lees-Edwards periodic boundary conditions and the SLLOD
algorithm~\cite{Evans90}.  In addition, the temperature
$T_{LJ}=0.1\,\varepsilon/k_B$ was maintained by the dissipative
particle dynamics (DPD) thermostat, which ensures that the particle
dynamics is not coupled to the imposed flow
profile~\cite{Soddemann03}. In the present study, the oscillation
period was fixed to $T=5000\,\tau$ and the strain amplitude was
varied in the range $0.03 \leqslant \gamma_0 \leqslant 0.07$.  The
data for the shear stress, potential energy, and atom positions were
collected during $600$ shear cycles for each value of the strain
amplitude.  The postprocessing analysis was performed only in one
sample due to computational restrictions.

\section{Results}
\label{sec:Results}


The structure and properties of metallic glasses depend strongly on
the details of the production and processing routes~\cite{Greer16}.
For example, it is well known that upon slower cooling, the glassy
system can reach states with lower potential energy and smaller
volume~\cite{Greer16}.  Moreover, under applied deformation, the
system can further explore different regions of the potential energy
landscape that are hardly accessible otherwise~\cite{Lacks04}. Thus,
it was demonstrated that one shear cycle with large strain amplitude
rejuvenates the glass by moving it to a state with shallower energy
minima, whereas one small-strain cycle overages the glass by
reaching a state with deeper energy minima~\cite{Lacks04}.
Furthermore, it was recently shown that with increasing cooling
rate, the strain-induced energy loss per strain (caused by particle
rearrangements) increases and glasses become more ductile and less
reversible~\cite{OHern17}.   In the present study, the binary
mixture in a high-temperature liquid state is instantaneously
quenched across the glass transition and then subjected to periodic
deformation for hundreds of cycles with strain amplitudes above and
below the yielding transition.

\vskip 0.05in

\begin{figure}[t]
\includegraphics[width=12.cm,angle=0]{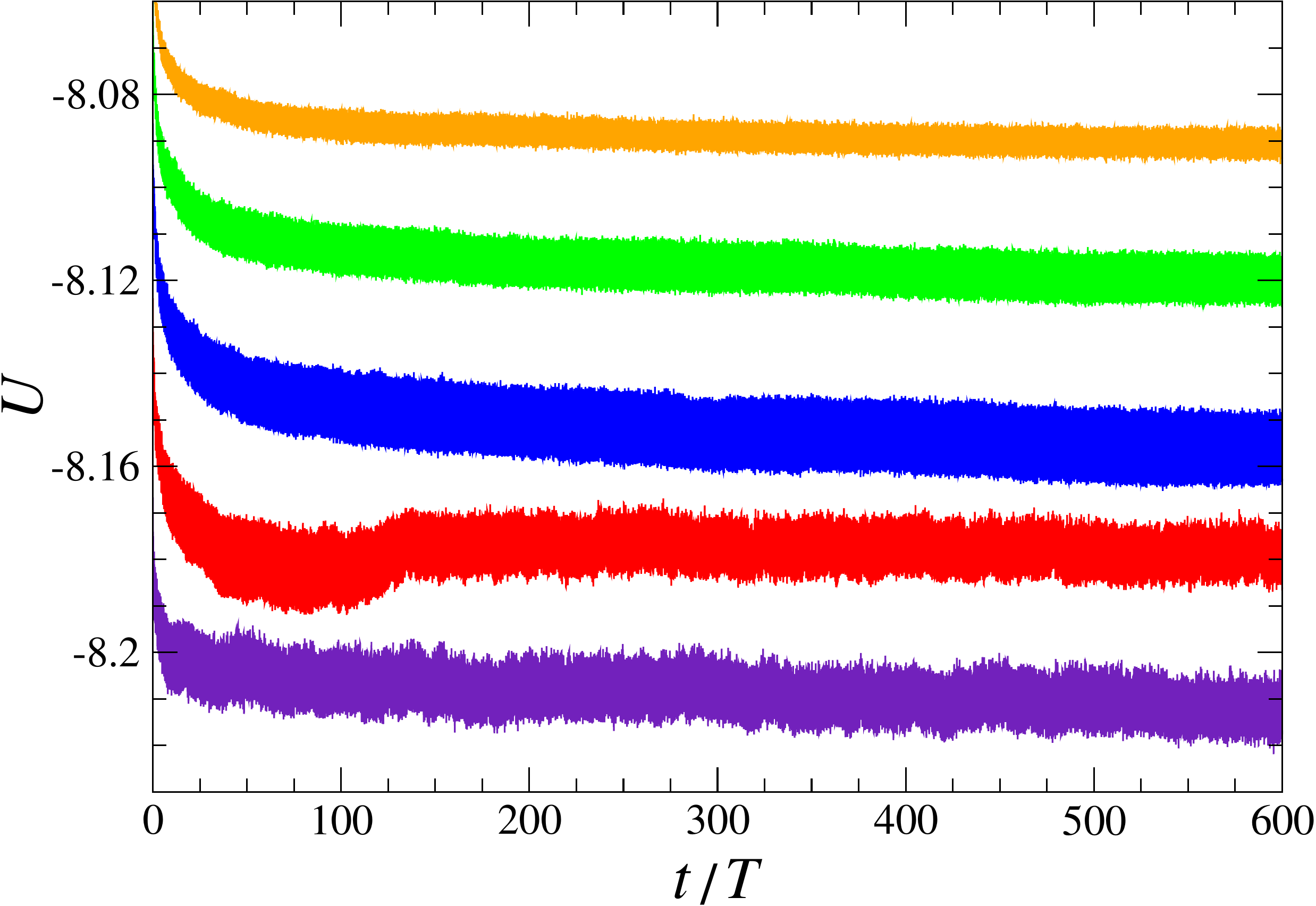}
\caption{(Color online) The time dependence of the potential energy
per particle $U$ (in units of $\varepsilon$) during 600 shear cycles
for the strain amplitudes $\gamma_{0} = 0.03$ (orange), $0.04$
(green), $0.05$ (blue), $0.06$ (red), and $0.07$ (indigo). For
clarity, the data were displaced vertically by $+0.05\,\varepsilon$
for $\gamma_{0} = 0.03$, by $+0.03\,\varepsilon$ for $\gamma_{0} =
0.04$, by $-0.04\,\varepsilon$ for $\gamma_{0} = 0.06$, and by
$-0.08\,\varepsilon$ for $\gamma_{0} = 0.07$. The data for
$\gamma_{0} = 0.05$ were left as is. The oscillation period is
$T=5000\,\tau$ and the temperature is $T_{LJ}=0.1\,\varepsilon/k_B$.
}
\label{fig:poten_time_Tr1}
\end{figure}

The variation of the potential energy, $U/\varepsilon$, during $600$
shear cycles after the thermal quench to
$T_{LJ}=0.1\,\varepsilon/k_B$ is shown in
Fig.\,\ref{fig:poten_time_Tr1} for the strain amplitudes
$\gamma_0=0.03$, $0.04$, $0.05$, $0.06$ and $0.07$.   Note that the
data for all strain amplitudes except for $\gamma_0=0.05$ are
displaced vertically for clarity (see caption to
Fig.\,\ref{fig:poten_time_Tr1}).   It can be clearly seen that the
potential energy decreases rapidly during the first few tens of
shear cycles and then it gradually saturates to a nearly constant
value for each strain amplitude.   In the case $\gamma_0=0.06$,
however, a markedly different behavior is observed; namely, the
local minimum is developed at $t\approx80\,T$ (see
Fig.\,\ref{fig:poten_time_Tr1}). In other words, the system is first
driven to a relatively deep energy minimum, where the amplitude of
energy oscillations is enhanced, followed by a crossover to a steady
state with a higher potential energy and reduced energy amplitude.
For all cases presented in Fig.\,\ref{fig:poten_time_Tr1}, the
steady state with the lowest potential energy, $U\approx
-8.16\,\varepsilon$, is achieved at $\gamma_0=0.05$, which suggests
that the critical strain amplitude for the yielding transition is in
the range between $0.05$ and $0.06$ at the temperature
$T_{LJ}=0.1\,\varepsilon/k_B$.  For comparison, the cyclic loading
with $\gamma_0=0.05$ at lower temperatures,
$T_{LJ}\leqslant10^{-2}\,\varepsilon/k_B$, resulted in the states
with the potential energy $U\approx -8.26\,\varepsilon$, and the
critical strain amplitude was found to be greater than
$0.07$~\cite{Priezjev18}. We also comment that the potential energy
saturates at $U\approx -8.12\,\varepsilon$ when the system is
instantaneously quenched to $T_{LJ}=0.1\,\varepsilon/k_B$ and
evolves in the absence of periodic shear.


\vskip 0.05in

\begin{figure}[t]
\includegraphics[width=12.cm,angle=0]{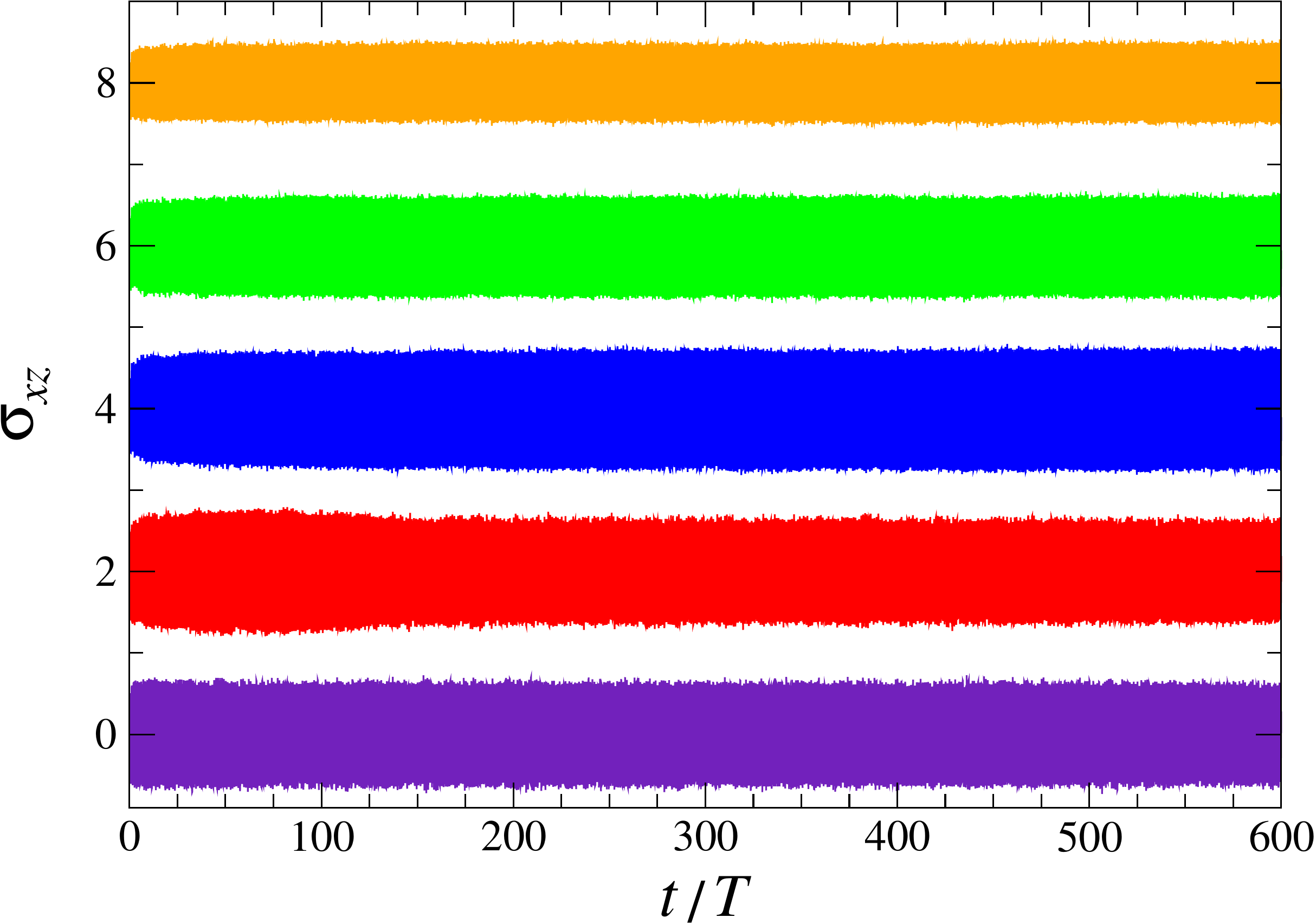}
\caption{(Color online) The variation of the shear stress
$\sigma_{xz}$ (in units of $\varepsilon\sigma^{-3}$) for the strain
amplitudes $\gamma_{0} = 0.03$ (orange), $0.04$ (green), $0.05$
(blue), $0.06$ (red), and $0.07$ (indigo). The data were displaced
by $8.0\,\varepsilon\sigma^{-3}$ for $\gamma_{0} = 0.03$, by
$6.0\,\varepsilon\sigma^{-3}$ for $\gamma_{0} = 0.04$, by
$4.0\,\varepsilon\sigma^{-3}$ for $\gamma_{0} = 0.05$, and by
$2.0\,\varepsilon\sigma^{-3}$ for $\gamma_{0} = 0.06$. The period of
oscillation is $T=5000\,\tau$. }
\label{fig:stress_time_Tr1}
\end{figure}

The time dependence of the shear stress during $600$ cycles is
illustrated in Fig.\,\ref{fig:stress_time_Tr1} for the same strain
amplitudes $\gamma_0=0.03$, $0.04$, $0.05$, $0.06$ and $0.07$ as in
Fig.\,\ref{fig:poten_time_Tr1}. Notice that the mean value of shear
stress is zero but the data for different strain amplitudes in
Fig.\,\ref{fig:stress_time_Tr1} are displaced upward for
visualization. It can be observed that the amplitude of stress
oscillations in steady state increases when the strain amplitude is
varied from $\gamma_0=0.03$ to $0.05$.  In the case $\gamma_0=0.06$,
the stress amplitude during the first 100 cycles becomes even larger
than for $\gamma_0=0.05$ but upon approaching steady state,
$t\gtrsim 200\,T$, the shear stress is reduced, indicating
significant plastic deformation in the material (discussed below).
This behavior is consistent with the appearance of a shallow minimum
in the potential energy for the strain amplitude $\gamma_0=0.06$
reported in Fig.\,\ref{fig:poten_time_Tr1}.  Finally, the transient
regime to steady state is reduced to only a few cycles for the
strain amplitude $\gamma_0=0.07$, and the stress amplitudes become
nearly the same for the cases $\gamma_0=0.06$ and $0.07$ when
$t\gtrsim 200\,T$.

\vskip 0.05in

\begin{figure}[t]
\includegraphics[width=12.cm,angle=0]{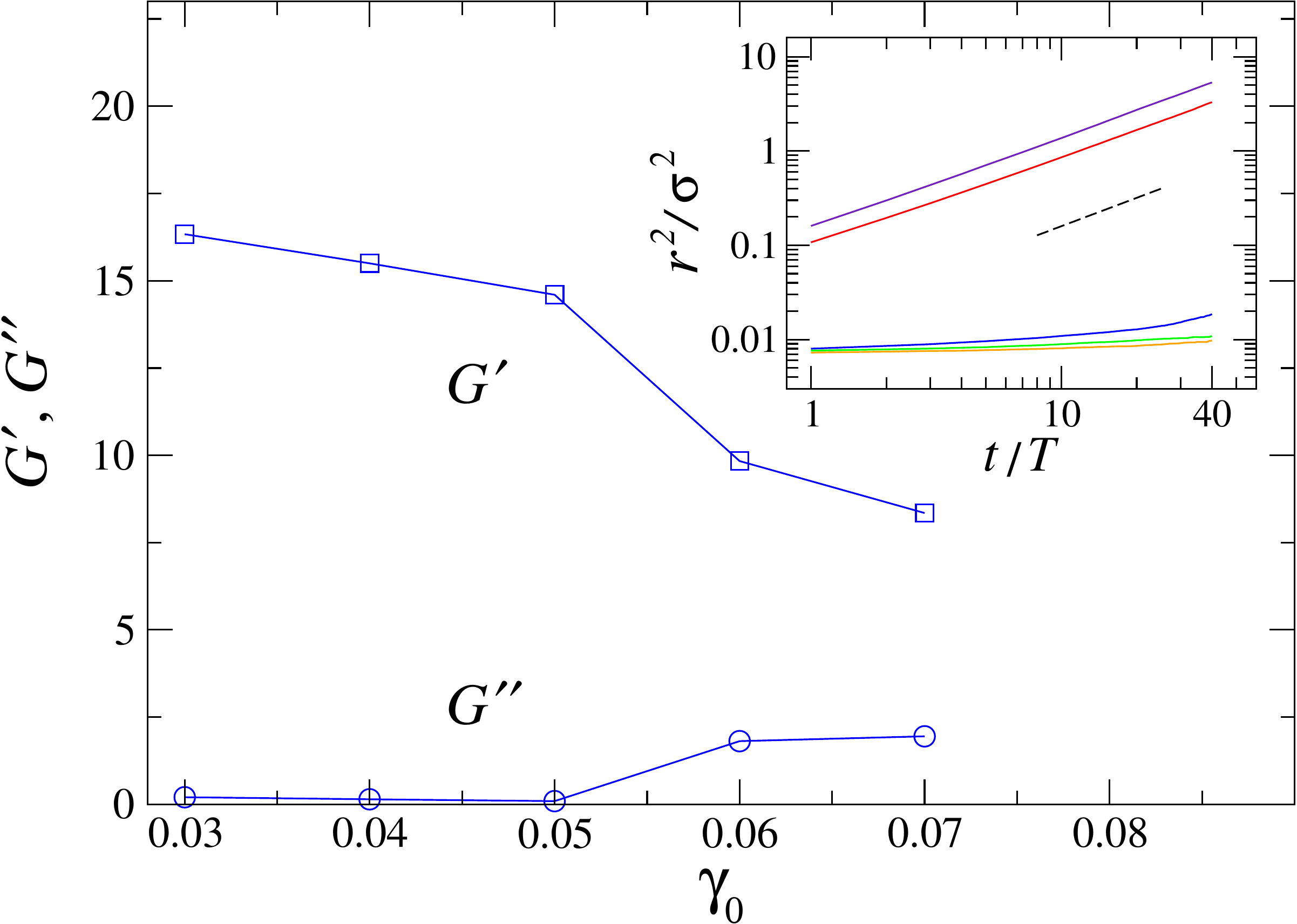}
\caption{(Color online) The storage ($G^{\prime}$) and loss
($G^{\prime\prime}$) moduli (in units of $\varepsilon\sigma^{-3}$)
as a function of the strain amplitude.   The inset shows the mean
square displacement of atoms for the strain amplitudes $\gamma_{0} =
0.03$ (orange), $0.04$ (green), $0.05$ (blue), $0.06$ (red), and
$0.07$ (indigo). The data were averaged over the last $40$ shear
cycles. The straight dashed line indicates the unit slope. }
\label{fig:moduli_msd}
\end{figure}

We next plot the storage ($G^{\prime}$) and loss
($G^{\prime\prime}$) moduli as a function of the strain amplitude in
Fig.\,\ref{fig:moduli_msd}. The data were computed from the shear
stress curves, $\sigma_{xz}(t)$, in steady state using the
definitions
$G^{\prime}=\sigma^{max}_{xz}/\gamma_0\,\text{cos}(\delta)$ and
$G^{\prime\prime}=\sigma^{max}_{xz}/\gamma_0\,\text{sin}(\delta)$,
where $\delta$ is the phase difference between stress and
strain~\cite{McKinley11}.  As is evident, the storage modulus is
much larger than the loss modulus at small strain amplitudes,
$\gamma_0\leqslant0.05$. With further increasing strain amplitude,
$\gamma_0\geqslant0.06$, the difference between the storage and loss
moduli significantly decreases, indicating the onset of energy
dissipation due to plastic deformation.  Along with the mechanical
properties, we present the mean square displacement of atoms for
different strain amplitudes in the inset to
Fig.\,\ref{fig:moduli_msd}.  It is clearly seen that a transition
between nearly reversible dynamics and diffusive behavior occurs at
the strain amplitude $\gamma_0=0.06$. These results are consistent
with conclusions of the previous study of jammed solids subjected to
large-amplitude oscillatory shear, where it was demonstrated that
the critical strain amplitude associated with the onset of particle
diffusion is smaller than the strain amplitude at which a crossing
of $G^{\prime}$ and $G^{\prime\prime}$ occurs~\cite{Kawasaki16}.  We
finally comment that a more gradual increase of the mean square
displacement curve at the strain amplitude $\gamma_0=0.06$ reported
in the previous study~\cite{Priezjev13} is due to much smaller
oscillation period $T\approx314\,\tau$ rather than $T=5000\,\tau$
used in the present study.

\vskip 0.05in


A complementary analysis of atomic rearrangements during oscillatory
shear deformation involves the so-called nonaffine displacements of
atoms, which are defined as a deviation from a local linear
transformation~\cite{Falk98}. In practice, the nonaffine measure can
be estimated numerically using the transformation matrix
$\mathbf{J}_i$, which maps all vectors between the $i$-th atom and
its neighbors during the time interval $\Delta t$ as follows:
\begin{equation}
D^2(t, \Delta t)=\frac{1}{N_i}\sum_{j=1}^{N_i}\Big\{
\mathbf{r}_{j}(t+\Delta t)-\mathbf{r}_{i}(t+\Delta t)-\mathbf{J}_i
\big[ \mathbf{r}_{j}(t) - \mathbf{r}_{i}(t)    \big] \Big\}^2,
\label{Eq:D2min}
\end{equation}
where the sum is carried over the neighbors within the cutoff
distance of $1.5\,\sigma$ from $\mathbf{r}_{i}(t)$. It was recently
demonstrated that in periodically sheared glasses, a large fraction
of atoms undergo repetitive nonaffine displacements with amplitudes
that are broadly distributed~\cite{Priezjev16,Priezjev16a}.  Above
the yield strain, some atoms in well-annealed (slowly quenched)
binary glasses start to rearrange irreversibly, and, after a number
of transient shear cycles, the atoms with large nonaffine
displacements were shown to organize into a system-spanning shear
band, whose thickness increases at larger strain
amplitudes~\cite{Priezjev17}.

\vskip 0.05in

\begin{figure}[t]
\includegraphics[width=12.cm,angle=0]{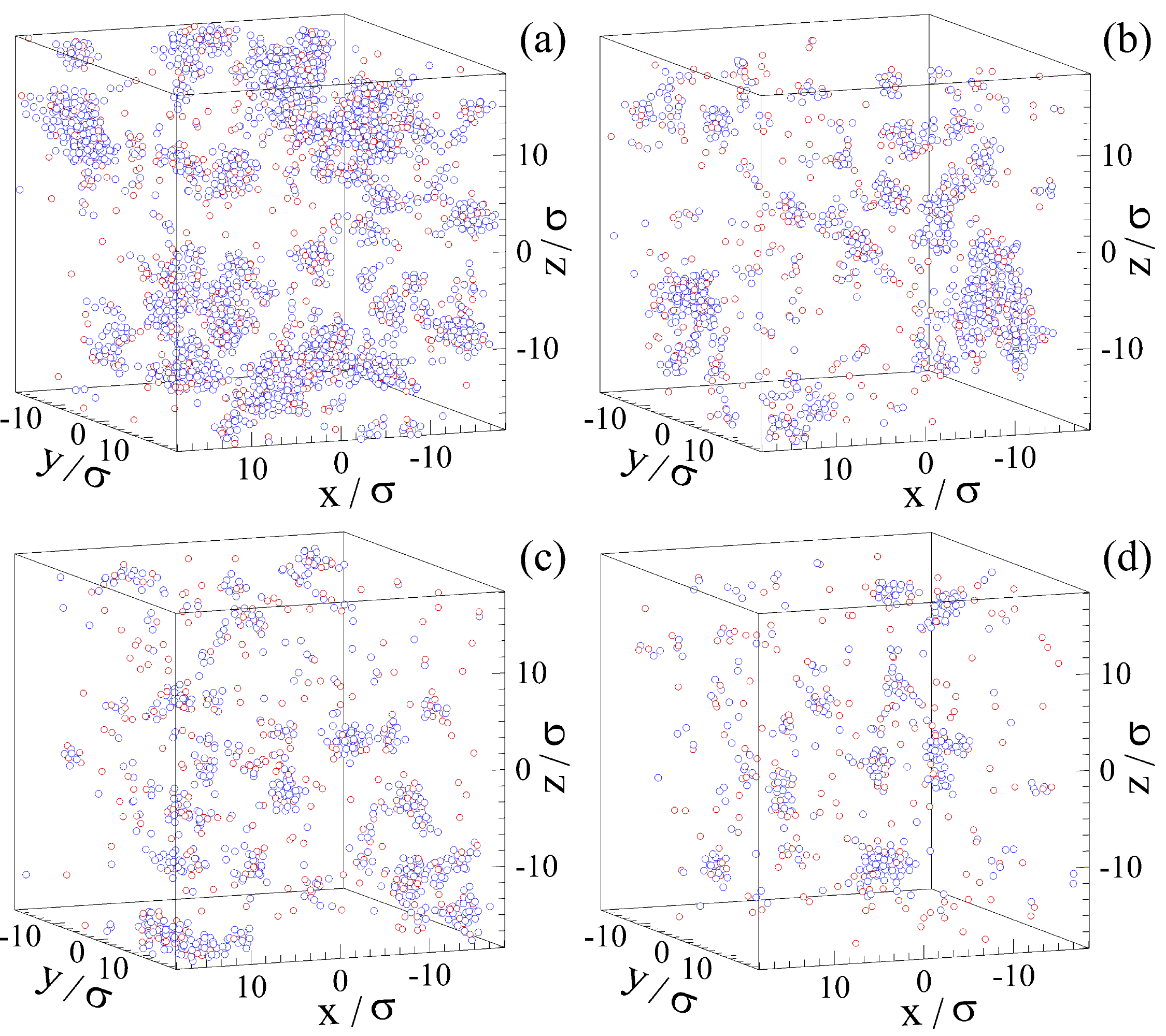}
\caption{(Color online) Snapshots of atomic configurations for the
strain amplitude $\gamma_{0}=0.03$, temperature
$T_{LJ}=0.1\,\varepsilon/k_B$, and nonaffine measure (a)
$D^2(19T,T)>0.04\,\sigma^2$, (b) $D^2(79T,T)>0.04\,\sigma^2$, (c)
$D^2(199T,T)>0.04\,\sigma^2$, and (d) $D^2(599T,T)>0.04\,\sigma^2$.}
\label{fig:snapshot_clusters_gam03_Tr1}
\end{figure}
\begin{figure}[t]
\includegraphics[width=12.cm,angle=0]{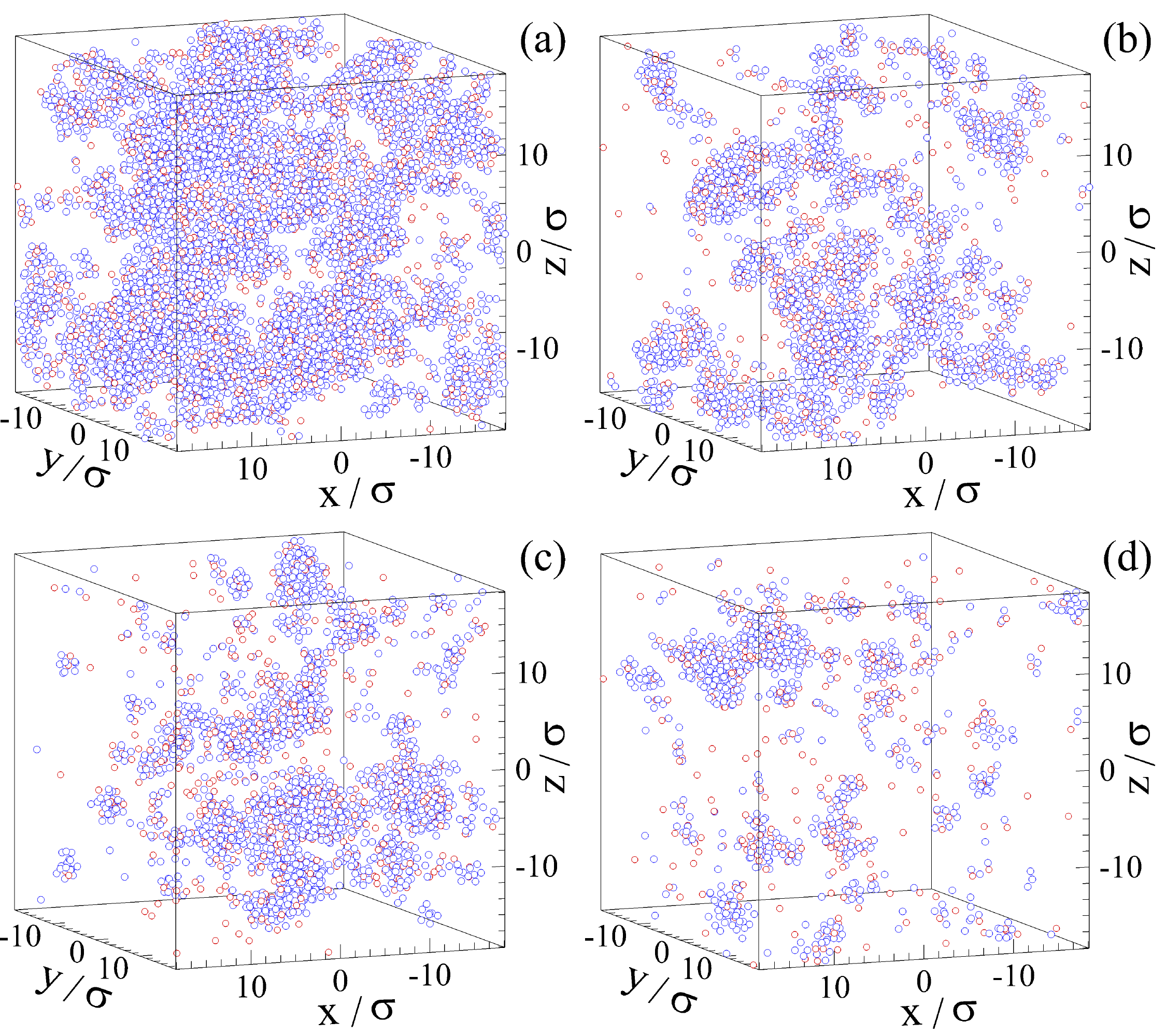}
\caption{(Color online) Spatial configurations of atoms with large
nonaffine displacements (a) $D^2(19T,T)>0.04\,\sigma^2$, (b)
$D^2(79T,T)>0.04\,\sigma^2$, (c) $D^2(199T,T)>0.04\,\sigma^2$, and
(d) $D^2(599T,T)>0.04\,\sigma^2$. The strain amplitude is
$\gamma_{0}=0.05$ and temperature is $T_{LJ}=0.1\,\varepsilon/k_B$.}
\label{fig:snapshot_clusters_gam05_Tr1}
\end{figure}
\begin{figure}[t]
\includegraphics[width=12.cm,angle=0]{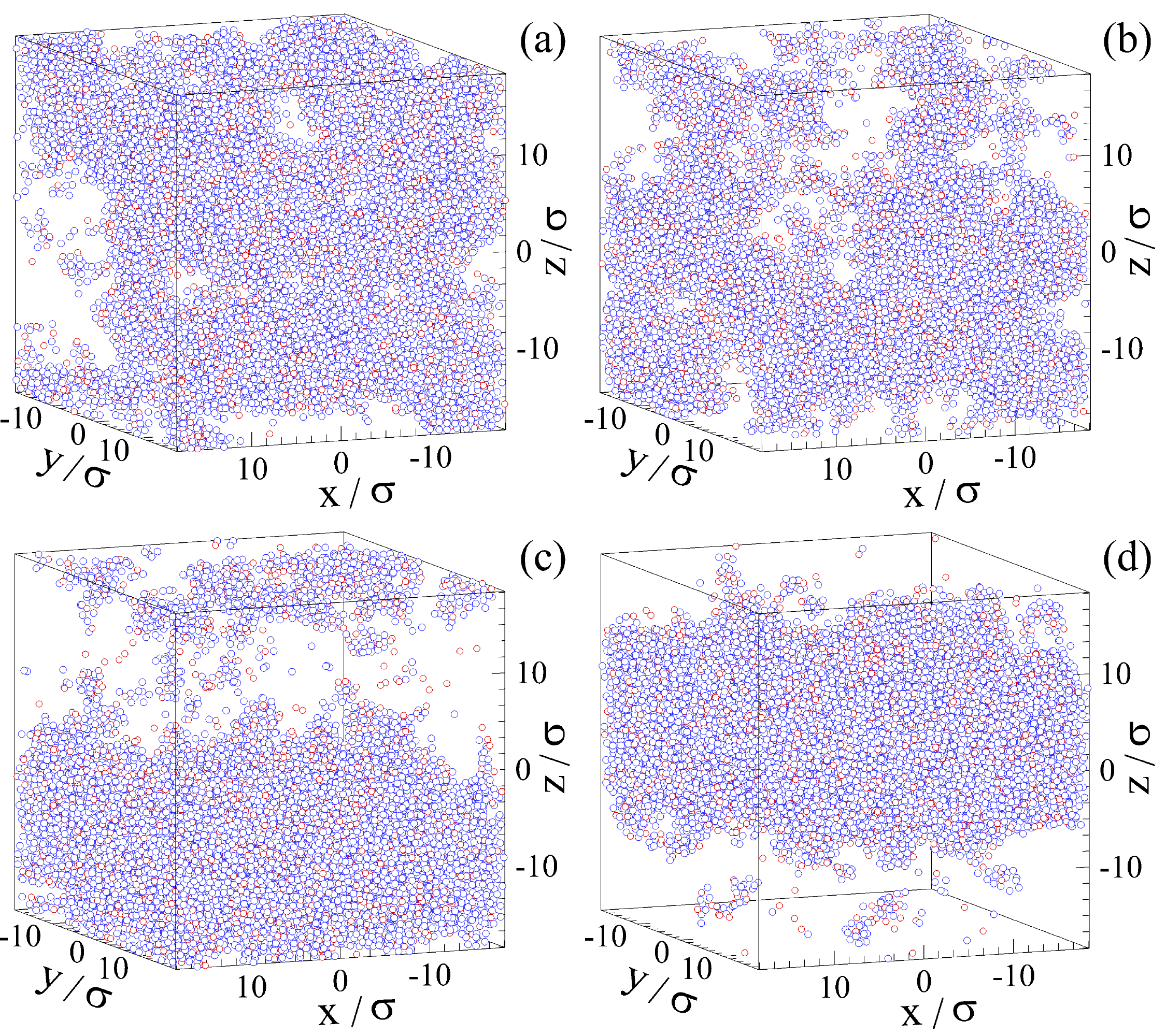}
\caption{(Color online) Atomic positions for the strain amplitude
$\gamma_{0}=0.06$, temperature $T_{LJ}=0.1\,\varepsilon/k_B$, and
nonaffine measure (a) $D^2(19T,T)>0.04\,\sigma^2$, (b)
$D^2(79T,T)>0.04\,\sigma^2$, (c) $D^2(199T,T)>0.04\,\sigma^2$, and
(d) $D^2(599T,T)>0.04\,\sigma^2$.}
\label{fig:snapshot_clusters_gam06_Tr1}
\end{figure}
\begin{figure}[t]
\includegraphics[width=12.cm,angle=0]{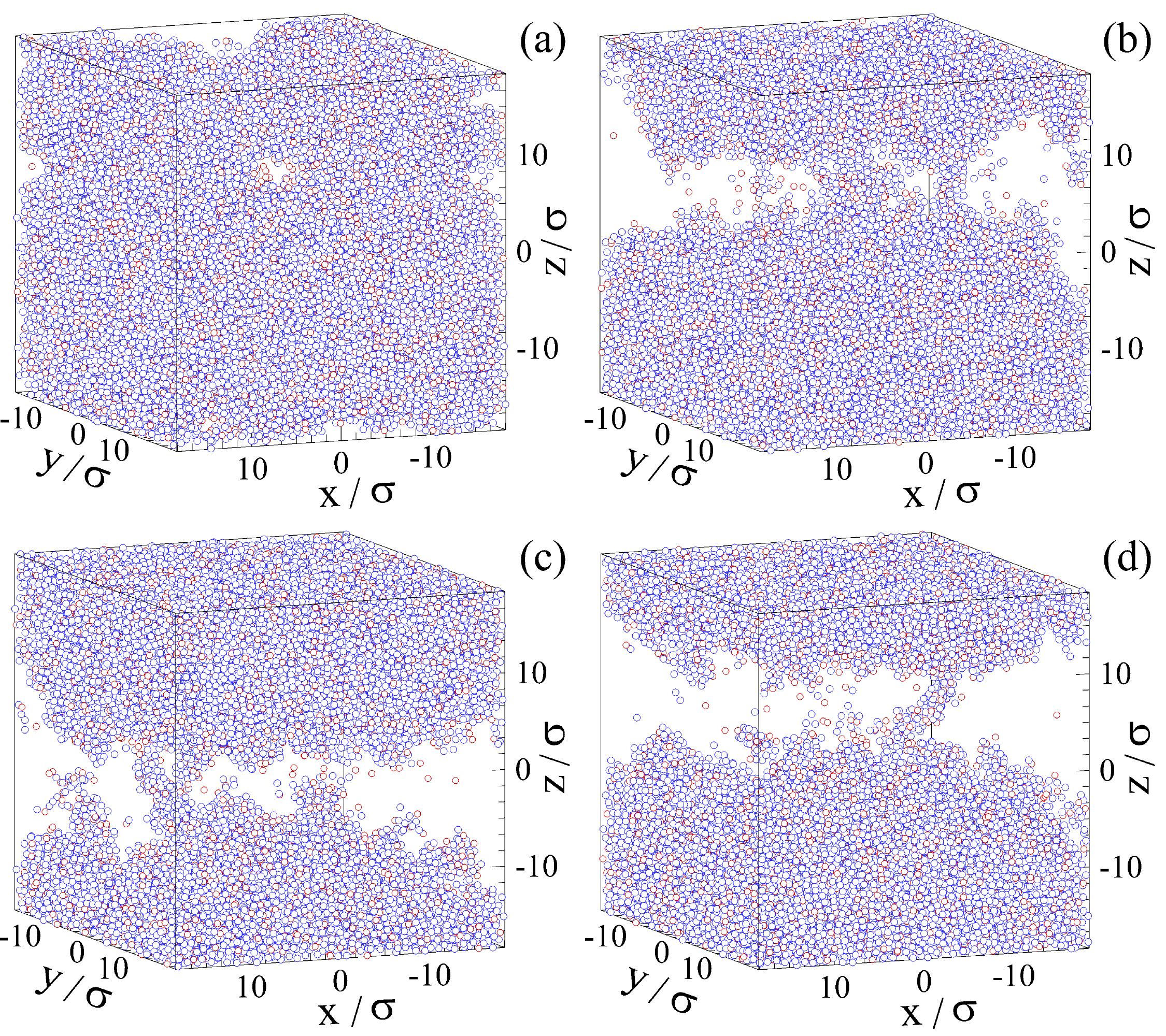}
\caption{(Color online) Snapshots of atomic positions for the strain
amplitude $\gamma_{0}=0.07$ and nonaffine measure (a)
$D^2(19T,T)>0.04\,\sigma^2$, (b) $D^2(79T,T)>0.04\,\sigma^2$, (c)
$D^2(199T,T)>0.04\,\sigma^2$, and (d) $D^2(599T,T)>0.04\,\sigma^2$.
The temperature of the system is $T_{LJ}=0.1\,\varepsilon/k_B$. }
\label{fig:snapshot_clusters_gam07_Tr1}
\end{figure}

Spatial configurations of atoms with large nonaffine displacements
are displayed in Figs.\,\ref{fig:snapshot_clusters_gam03_Tr1},
\ref{fig:snapshot_clusters_gam05_Tr1},
\ref{fig:snapshot_clusters_gam06_Tr1}, and
\ref{fig:snapshot_clusters_gam07_Tr1} for the strain amplitudes
$\gamma_0=0.03$, $0.05$, $0.06$ and $0.07$, respectively.   It can
be seen in Figs.\,\ref{fig:snapshot_clusters_gam03_Tr1} and
\ref{fig:snapshot_clusters_gam05_Tr1} that at small strain
amplitudes, $\gamma_0 \leqslant 0.05$, the atoms with
$D^2>0.04\,\sigma^2$ form relatively large clusters during $20$-th
cycle, and upon further annealing, the number of atoms with nearly
reversible trajectories increases.  This behavior is consistent with
the gradual decay of the potential energy reported in
Fig.\,\ref{fig:poten_time_Tr1} and large values of the storage
modulus in Fig.\,\ref{fig:moduli_msd} for $\gamma_0 \leqslant 0.05$.
In sharp contrast, at large strain amplitudes, $\gamma_0 \geqslant
0.06$, most of the atoms initially undergo large
($D^2>0.04\,\sigma^2$) nonaffine displacements, and, after a number
of transient shear cycles, a system-spanning shear band is formed
(see Figs.\,\ref{fig:snapshot_clusters_gam06_Tr1} and
\ref{fig:snapshot_clusters_gam07_Tr1}).    Notice that in both
cases, $\gamma_0 = 0.06$ and $0.07$, the location of the shear band
is displaced along the $\hat{z}$ direction over consecutive cycles.
The appearance of the shear band at large strain amplitudes
correlates well with the onset of dissipation (larger loss modulus)
and enhanced diffusion reported in Fig.\,\ref{fig:moduli_msd}. We
further remark that cycling loading of poorly annealed glasses at
lower temperatures, $T_{LJ}\leqslant10^{-2}\,\varepsilon/k_B$,
during $600$ shear cycles did not result in the formation of shear
bands for the strain amplitudes $0.03 \leqslant \gamma_0 \leqslant
0.07$~\cite{Priezjev18}.

\vskip 0.05in

\begin{figure}[t]
\includegraphics[width=12.cm,angle=0]{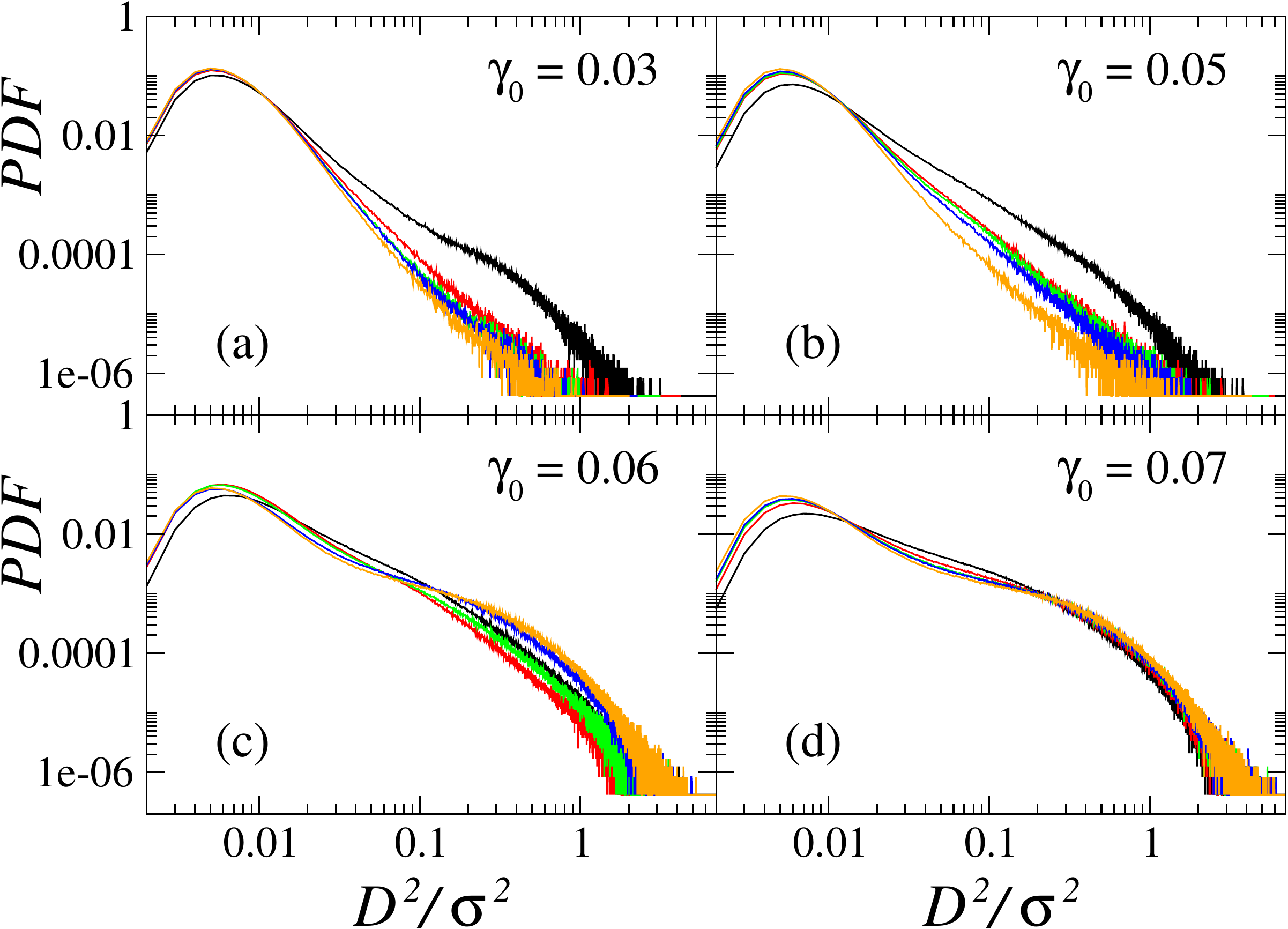}
\caption{(Color online) The normalized probability distribution
functions of the nonaffine measure $D^2(t,T)$ for the strain
amplitudes (a) $\gamma_{0}=0.03$, (b) $\gamma_{0}=0.05$, (c)
$\gamma_{0}=0.06$, and (d) $\gamma_{0}=0.07$. The data are averaged
during the following time intervals $0\leqslant t\leqslant40\,T$
(black curves), $40\,T\leqslant t\leqslant80\,T$ (red curves),
$80\,T\leqslant t\leqslant120\,T$ (green curves), $160\,T\leqslant
t\leqslant200\,T$ (blue curves), and $560\,T\leqslant
t\leqslant600\,T$ (orange curves). }
\label{fig:pdf_D2min}
\end{figure}

The observed sequence of patterns (shear bands vs. disconnected
clusters) of atoms with large nonaffine displacements is reflected
in the shape of probability distribution functions of the nonaffine
measure.  Next, the probability distribution function of $D^2(t,T)$,
averaged within narrow time intervals, are presented in
Fig.\,\ref{fig:pdf_D2min} for the strain amplitudes $\gamma_0=0.03$,
$0.05$, $0.06$ and $0.07$.  In agreement with the previous
studies~\cite{Schall12,Priezjev16,Priezjev16a,Priezjev17}, the
nonaffine displacements for all strain amplitudes and time intervals
are broadly distributed (see Fig.\,\ref{fig:pdf_D2min}).  However,
the time dependence of the probability distributions is markedly
different for $\gamma_0 \leqslant 0.05$ and $\gamma_0 \geqslant
0.06$.  In particular, the shape of the probability distributions
for $\gamma_0 = 0.03$ and $0.05$ in
Fig.\,\ref{fig:pdf_D2min}\,(a,\,b) becomes more narrow over
consecutive cycles, indicating progressively more reversible
dynamics; while for $\gamma_0 = 0.07$ shown in
Fig.\,\ref{fig:pdf_D2min}\,(d), the opposite trend emerges.  The
most peculiar case is shown for the strain amplitude $\gamma_0=0.06$
in Fig.\,\ref{fig:pdf_D2min}\,(c), where the distribution of $D^2$
first becomes more narrow and then slightly broadens as the number
of cycles increases.    This behavior is correlated with the
nonmonotonic transient of the potential energy shown in
Fig.\,\ref{fig:poten_time_Tr1} and the temporary increase of the
shear stress oscillations in Fig.\,\ref{fig:stress_time_Tr1}.
Overall, we conclude that at the higher target temperature
$T_{LJ}=0.1\,\varepsilon/k_B$ (in comparison with the range
$10^{-6}\,\varepsilon/k_B\leqslant
T_{LJ}\leqslant10^{-2}\,\varepsilon/k_B$ considered in the previous
study~\cite{Priezjev18}), the critical strain amplitude of the
yielding transition is decreased below $\gamma_0 = 0.06$.  We
finally comment that a decrease in storage modulus and proliferation
of large nonaffine particle displacements at the strain amplitude
$\gamma_0 = 0.06$ might be related to the onset of yielding
formulated in terms of loss of long-lived nearest
neighbors~\cite{ZacMex18}.

\section{Conclusions}

In summary, non-equilibrium molecular dynamics simulations were
performed to examine the dynamic response of a three-dimensional
model glass to oscillatory shear deformation. The model glass was
represented by a binary Lennard-Jones mixture that was rapidly
quenched from a high-temperature liquid state to a temperature of
about a quarter of the glass transition temperature.  Following the
thermal quench, the binary glass was subjected to periodic shear at
constant volume over hundreds of cycles. To ensure that particle
dynamics is not coupled to the imposed flow profile, the dissipative
particle dynamics thermostat was applied during periodic shear
deformation.

\vskip 0.05in

It was shown that with increasing strain amplitude up to a critical
value, the potential energy acquires progressively lower minima,
while the amplitude of stress oscillations in steady state becomes
larger. Moreover, the gradual decay of the potential energy is
associated with sparse clusters of atoms with large nonaffine
displacements.  Above the yielding transition, the viscoelastic
damping increases and the potential energy in steady state becomes
higher.  The initial stage of the structural relaxation process
involves transient clusters of atoms with large nonaffine
displacements, followed by the formation of a permanent shear band
that runs across the whole system.   Upon cyclic loading, the width
of the probability distribution function of nonaffine displacements
increases (decreases) for strain amplitudes above (below) the
yielding transition.

\section*{Acknowledgments}

Financial support from the National Science Foundation (CNS-1531923)
is gratefully acknowledged. The article was prepared within the
framework of the Basic Research Program at the National Research
University Higher School of Economics (HSE) and supported within the
framework of a subsidy by the Russian Academic Excellence Project
`5-100'. The molecular dynamics simulations were performed using the
LAMMPS numerical code developed at Sandia National
Laboratories~\cite{Lammps}. Computational work in support of this
research was performed at Michigan State University's High
Performance Computing Facility and the Ohio Supercomputer Center.


%

%

%


%

%

%

%

%

\bibliographystyle{prsty}

\end{document}